# Photoacoustic digital brain: numerical modelling and image reconstruction via deep learning

Tengbo Lyu, Jiadong Zhang, Zijian Gao, Changchun Yang, *Student Member, IEEE*, Feng Gao, and Fei Gao, *Member, IEEE*

*Abstract*—Photoacoustic tomography (PAT) is a newly developed medical imaging modality, which combines the advantages of pure optical imaging and ultrasound imaging, owning both high optical contrast and deep penetration depth. Very recently, PAT is studied in human brain imaging. Nevertheless, while ultrasound waves are passing through the human skull tissues, the strong acoustic attenuation and aberration will happen, which causes photoacoustic signals' distortion. In this work, we use 10 magnetic resonance angiography (MRA) human brain volumes, and manually segment them to obtain the 2D human brain numerical phantoms for PAT. The numerical phantoms contain six kinds of tissues which are scalp, skull, white matter, gray matter, blood vessel and cerebral cortex. For every numerical phantom, optical properties are assigned to every kind of tissues. Then, Monte-Carlo based optical simulation is deployed to obtain the photoacoustic initial pressure. Then, we made two k-wave simulation cases: one takes inhomogeneous medium and uneven sound velocity into consideration, and the other not. Then we use the sensor data of the former one as the input of U-net, and the sensor data of the latter one as the output of U-net to train the network. We randomly choose 7 human brain PA sinograms as the training dataset and 3 human brain PA sinograms as the testing set. The testing result shows that our method could correct the skull acoustic aberration and obtain the blood vessel distribution inside the human brain satisfactorily.

## I. Introduction

Cerebral vascular diseases have become one of the most dangerous killers in the world. Especially, stroke has ranked among the top ten pathogenies of death. Effective and timely detection of the cerebral vascular damaged position or oxygen supply can greatly reduce the mortality of patients. Clinically, there are three major diagnosis methods to detect cerebral vascular disease. They are magnetic resonance angiography (MRA) [1], computed tomography angiography (CTA) [2] and digital subtraction angiography (DSA) [3], which are illustrated in Fig 1.

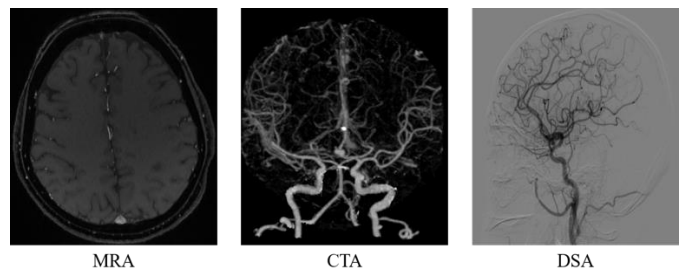

Fig. 1. Three major cerebral vascular disease diagnosis methods.

Though the above three methods can present the clear vascular structure, they have their own drawbacks and application limitations. The biggest drawback of MRA is its low imaging speed. Too long scanning time makes it less useful in diagnosis of acute patients. While CTA saves much more scanning time. However, exposing to radiation environment is still a big concern in real life. DSA is the golden standard of cerebral vascular diagnosis. But compared with non-destructive testing, it needs radiography operation to show the perfusion process. The cost of DSA is much more expensive than the other two. These limitations intrigue us to find new non-invasive and real-time imaging technologies.

As a hybrid imaging modality, photoacoustic (PA) imaging combines the advantages of both pure optical imaging (high optical contrast) and ultrasound imaging (deep penetration depth) [4]. Benefited from its vascular detection sensitivity and blood oxygen saturation quantification, PA imaging has become a promising non-invasive medical imaging modality in

Tengbo Lyu and Jiadong Zhang contributed equally to this work.

This research was funded by Start-up grant of ShanghaiTech University (F-0203-17-004), Natural Science Foundation of Shanghai (18ZR1425000), and National Natural Science Foundation of China (61805139).

The authors declare no conflicts of interest.

T. Lyu, J. Zhang, Z. Gao, C. Yang, F. Gao, F. Gao are all with Hybrid Imaging System Laboratory, Shanghai Engineering Research Center of Intelligent Vision and Imaging, School of Information Science and Technology, Shanghai Tech University, Shanghai 201210, China.

both preclinical and clinical application in recent decades. From now on, PA imaging has shown great potential in breast tumor classification, PA guided needle tip tracking, skin cancer detection and so on.

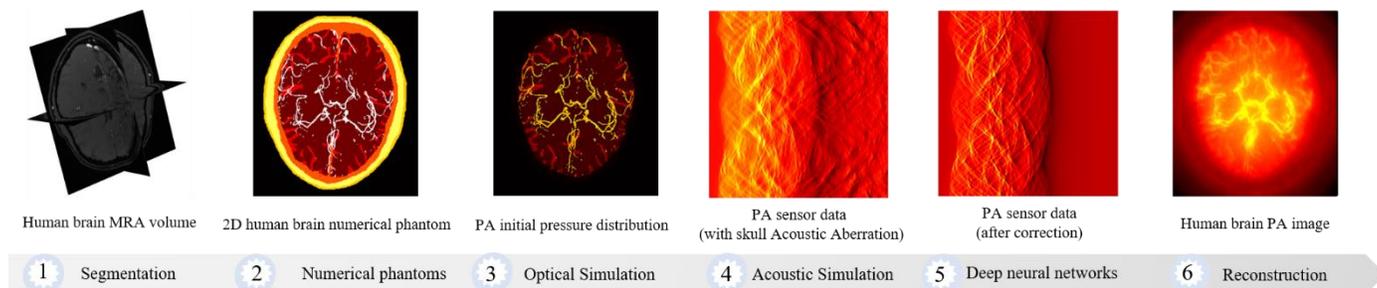

Fig. 2. The whole process of proposed method.

Considering that PA imaging plays an outstanding role in reveling vascular, it seems that PA imaging can be competent in the cerebral vascular diagnosis task. Liming Nie *et al* proved the feasibility that using a photon recycler can enable light to pass through the whole adult human skulls for the first time [5]. It starts the new era in human brain PA imaging. However, the aberration from wave reflection and refraction caused by skull tissue makes PA image reconstruction much difficult. In this regard, more works based on the abovementioned papers appear to correct aberration and make a big step towards the clinical PA human brain imaging. Chao Huang *et al* proposed to combine PAT with the skull composition and morphology from adjunct X-ray CT modality to correct aberration [6]. This work performs well in both phantoms and monkey head skull-induced distortions. Leila Mohammadi *et al* proposed a deterministic ray-tracing simulation framework [7]. Their model contains the attenuation and dispersion effects from wave reflection, refraction. They implement their work on the large 3D phantoms and get pretty good results with high computation speed. Shuai Na *et al* proposed to take skull heterogeneity into count to build a three-layer human brain model [8]. With wave reflection and refraction coefficients modified, the proposed layered universal back-projection can correct the acoustic aberration, then they prove the effectiveness on real *ex vivo* adult human skull with 64-element PAT system. Leila Mohammadi et al also chose to take attenuation and dispersion effect into a mathematical model [7]. They demonstrate the work's effectiveness on skull aberration correction in PA human brain imaging. They further designed a vector space similarity model to compensate the skull-induced acoustic distortion that also works well in numerical phantoms with photoacoustic microscopy (PAM) system.

All abovementioned works focus on building a mathematical model as accurate as possible to correct the influence caused by skull-distortion. However, there are still unknown detailed and latent physical mechanism that cannot be reflected in mathematical model. So, we want to learn to correct acoustic aberration in a data-driven way. Deep learning is a good data-driven way to mine the latent data relationship, and we use deep learning network (U-net) to correct distorted PA signals [9]. We list all the contributions of this paper as follow:

● We generate a realistic 2D human brain PA numerical phantom database for imaging and vascular disease diagnosis research, which is available on the website: https://dx.doi.org/10.21227/evyp-d384.
● We use deep learning method to correct the acoustic aberration. With recovered PA signals, we can get the skull-unaffected PA human brain images.
● We use numerical phantoms to prove the effectiveness of proposed method with pretty good imaging quality.

The reminder of this paper is organized as follow: First, we will introduce some related work in Section 2. Next, we make detailed description in Section 3. Then we show the experiment results in Section 4. Finally, we draw conclusion in Section 5.

## II. RELATED WORK

### A. Biological Tissue Numerical Phantoms

The well-designed numerical phantoms are important for performing a simulation study, when lacking of enough ground truth, especially for human biological tissue objects. Here we will summarize some biological tissue numerical phantoms generation work and list parts of their database website links.
(1) Human breast numerical phantoms: Yang Lou *et al* proposed to extract four parts (vessel, skin, fat and fibroglandular tissue) from contrast-enhanced magnetic resonance images (MRI) modality with threshold segmentation method [10]. But they only generate few 2D numerical phantoms from 50 patients. Based on this work, Maura Dantuma *et al* proposed to add breast tumor tissue from mice additionally [11]. Ant they extend the 2D breast numerical phantoms to 3D, which makes simulation more realistic. Then Yaxin Ma *et al* proposed to extract four parts (skin, fat, tumor and fibro glandular tissue) from mammography [12]. The first three tissues are manually segmented and the fibro glandular is segmented by deep neural network. This work generates large amounts of 2D breast numerical phantoms. But lacking breast vessels greatly limits the usefulness of the dataset. Later, Tao Han *et al* proposed to extract four parts (skin, adipose, tumor and fibro glandular tissue) from a series of 2D US B-scan slices with manual segmentation [13]. The reconstructed 3D breast numerical phantoms are generated, but lacking vessel tissue either. After analyzing all the work above, Youwei Bao *et al* proposed to generate a more realistic breast numerical phantoms including various kinds of breast types [14]. They use a software (VICTRE breast phantom) to generate a comprehensive breast model including skin, nipple, lactiferous duct, terminal duct lobular unit, interlobular gland tissue, fat, suspensory ligament, muscle, artery, and vein. It is the most useful and easy tool to implement human breast phantom database.

(2) Human skin numerical phantoms: Jun Q. Lu *et al* proposed to build two-layer skin model volumes (epidermis and underlying dermis) with random generation [15]. The structure is relatively simple but useful to optical properties analysis. Based on this, Tengbo Lyu *et al* proposed to extract vessel tissue from 3D lung CT scans with Frangi filter segmentation [16]. Then by embedding the vessel tissue to a three-layer cube (epidermis, dermis and hypodermis), they generate 3890 3D skin tissue numerical phantoms, which is available on the website: https://ieee-dataport.org/documents/3d-skin-tissue-vessel-models-medical-image-analysis.

(3) Human brain numerical phantoms: Kamyar Firouzi *et al* proposed to generate a numerical brain phantom from MRI and MRA [17]. They segmented six kinds of biological tissue (skull, white matter, gray matter, cerebrospinal fluid, edema, tumor) with 3DSlicer software package from MRI and segmented vascular tissue with level-set algorithm from MRA. It is suitable to investigate the human brain tumor diseases instead of cerebral vascular diseases. Later, Shuai Na *et al* proposed to generate 2D human brain numerical phantoms by embedding a small piece of vessel model to oval skull models, which are both manually designed. These phantoms are relatively simple and discard lots of other brain tissue, like white matter.

### B. Deep learning-based signal processing

Deep learning has achieved great success in signal processing (e.g. enhancement, denoising and recovery). Sreedevi Gutta *et al* proposed to use a standard five layer fully connection neural network to enhance the bandwidth of PA signals [18]. The proposed deep neural network (DNN) can compensate the bandwidth loss caused by ultrasound transducers. With the bandwidth enhanced PA signals, they get the improved reconstructed PA images without significant computation burden. Based on this work, Navchetan Awasthi *et al* proposed to use convolutional neural network (CNN) to enhancement the PA signals [19]. Combined with U-net (a well-known CNN structure) and proposed Elu activation function, the method can compensate both PA signal bandwidth and limited-view information loss. The results show that the method works well in PAT system with less root mean square error (RMSE) and higher signal-to-noise ratio (SNR) value. Ben Luijten *et al* proposed to use DNN to learn an optimal set of apodization weights, which can adaptively enhance the ultrasound signals [20]. With back projection methods, the proposed deep learning-based adaptive signal processing framework achieves data-efficiency and robustness in ultrasound signal enhancement and image reconstruction. Derek Allman *et al* proposed to use CNN to distinguish PA source from reflection artifacts with the sampled PA signals [21]. This work greatly decreases the artifact influence in PA images with high accuracy.

### III. METHOD

In this section, we will make detailed description about the proposed method. We divide it into five parts and they are human brain numerical phantoms generation, optical simulation, acoustic simulation, acoustic aberration correction using U-net and PA image reconstruction, which is illustrated in Fig. 2. Next, we will use five separated chapters to cover the whole pipeline.

### A. Human Brain Numerical Phantom Generation

In order to make the simulation study as real as possible, we want to generate the human brain numerical phantoms from other medical image modalities. Considering MRA can provide good tissue contrast with rich vessel information, we choose 3D MRA volume to generate the 2D human brain numerical phantoms. The MRA dataset could be accessed from the website: https://www.nitrc.org/projects/icbmmra/.

From MRA, we can see five parts clearly, which are skull, vessel, gray matter, white matter and cerebral cortex. By segmenting these five parts, we can get the photoacoustic numerical phantoms. However, the segmentation task is challenging, especially for such a small amount of data. Lacking labels or the limitation of the number of training sets make the segmentation based on deep learning is unavailable. The traditional segmentation algorithms also don't work well with the five tissue parts. So, we decide to combine traditional image processing with manual segmentation. In MRA, vessels have higher value compared with other tissues caused by contrast agent, which makes vessels can be easily segmented with threshold or filter methods. After comparing the segmentation results, we choose 3D Frangi filter to segment vessel part [22]. Then we will manually segment other tissue. Because the manual segmentation process of the 3D volume (288 x 320 x 224) is time consuming, we only choose an upper slice (288 x 320) to generate 2D numerical phantoms. Considering the scalp is also an important part in human brain simulation, we additionally add a layer of scalp outside the skull. Then we combine the 2D projection of 3D vessel information with other tissue part to get the final human brain numerical phantoms. We use numbers 1-6 to denotes the six types of tissue. Number 1 denotes white matter, 2 denotes gray matter, 3 denotes cerebral cortex, 4 denotes scalp, 5 denotes skull and 6 denotes vessel. The whole process is illustrated in Fig. 3.

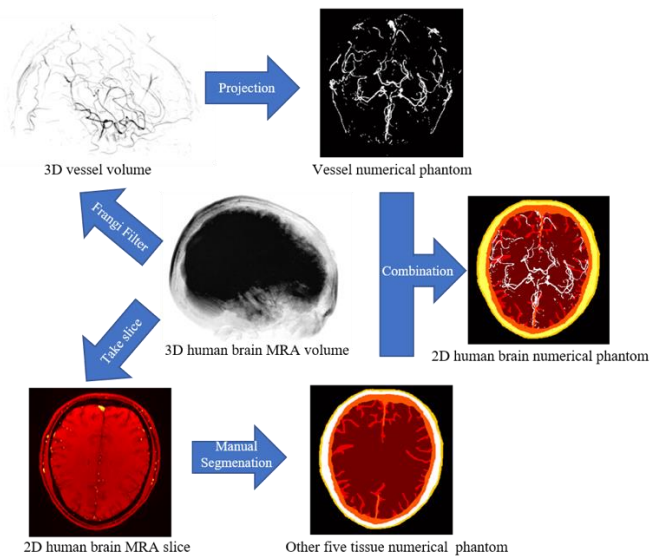

Fig. 3. 2D human brain numerical phantom generation process.

Considering light has to pass through some tissue from top of the head, we build pseudo-3D human brain numerical phantoms to make the optical simulation more realistic. Pseudo-3D means the numerical phantom is not generated from the real 3D MRA source. It is just spliced with several 2D numerical phantom slices. Illustrated in Fig 4, the laser has to go through scalp layer (Layer 1: only with scalp tissue), skull layer (Layer 2: with scalp and skull tissue), cerebral cortex layer (Layer 3: with scalp, skull and cerebral cortex tissue) to reach the final layer (Layer 4: with all tissue). Layer 4 is the human brain numerical phantom in Fig 3. Because different tissue has different size, we build the pseudo-3D human brain numerical phantom sizes 288 x 320 x 12 (2 Layer 1 on the top, 5 Layer 2 under Layer 1, 4 Layer 3 under Layer 2 and 1 Layer 4 at the bottom).

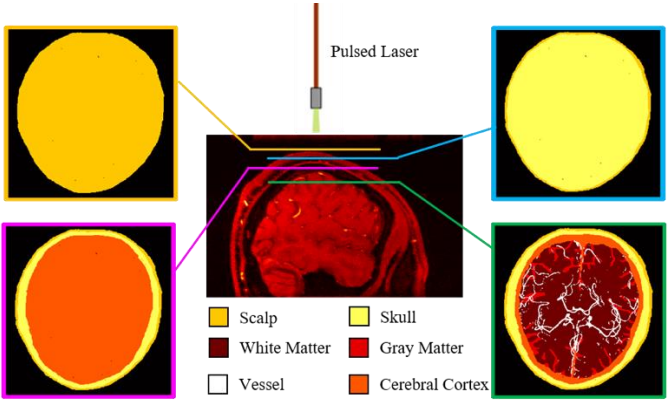

Fig. 4. Four kinds of layers and their positions from the pseudo-3D human brain numerical phantoms.

### B. Optical Simulation

After generating 10 human brain numerical phantoms, in this section, we will introduce the optical simulation. The optical fluence simulation is deployed using the 3D Monte Carlo method [23]. The optical and acoustic properties of the six kinds of tissues are shown in Table 1, which mainly include the optical absorption coefficients ($\mu_a$), the reduced scattering coefficients ($\mu_s$), anisotropy ($n$), refractive index ($g$), the sound speed ($c$) and the medium density ($d$). We use the MATLAB package MCXLAB to simulate the photons propagation inside the human brain numerical phantoms illuminated by the pulsed laser.

Without loss of generality, our optical simulation setup based on 3D Monte Carlo method is as follows: 1) the shape of the light source is planar; 2) the total number of photons to be simulated is 1 trillion; 3) the position and incident vector of the source is set properly to cover all the regions of interest; 4) the starting time and the ending time of the simulation is set to 0 and 1 nanoseconds respectively, and the time-gate width of the simulation is set to 0.01 nanoseconds. According to Table 1, given the pulsed laser excitation, the optical fluence through the six kinds of human brain tissues is computed and obtained.
This optical simulation is run on a Linux server with 4 GTX 1080Ti GPUs. It takes about tens of seconds to finish the photons propagation process. For the 10 human brain numerical phantoms, we obtain 10 optical fluence maps, the size of each optical fluence map is about 450 MB. Nine optical fluence maps of the human brain numerical phantoms are shown in Fig. 5. And finally, since our human brain numerical phantoms are pseudo 3D, we only need to slice the 12th slice of this 3D optical fluence volume for the following acoustic simulation.

Table 1. Optical and acoustic properties of the 2D human brain numerical phantom.

| | Optical properties | | | | Acoustic properties | |
|---|---|---|---|---|---|---|
| | $\mu_a$ | $\mu_s$ | g | n | c | d |
| White matter | 0.14 | 910 | 0.9 | 1.37 | 1500 | 1000 |
| Gray matter | 0.36 | 220 | 0.9 | 1.37 | 1500 | 1000 |
| Cerebral cortex | 0.06 | 2.4 | 0.9 | 1.43 | 1500 | 1000 |
| scalp | 0.18 | 190 | 0.9 | 1.37 | 1500 | 1000 |
| Skull | 0.16 | 160 | 0.9 | 1.43 | 3500 | 1900 |
| Vessel | 2.33 | 500 | 0.99 | 1.4 | 1500 | 1000 |

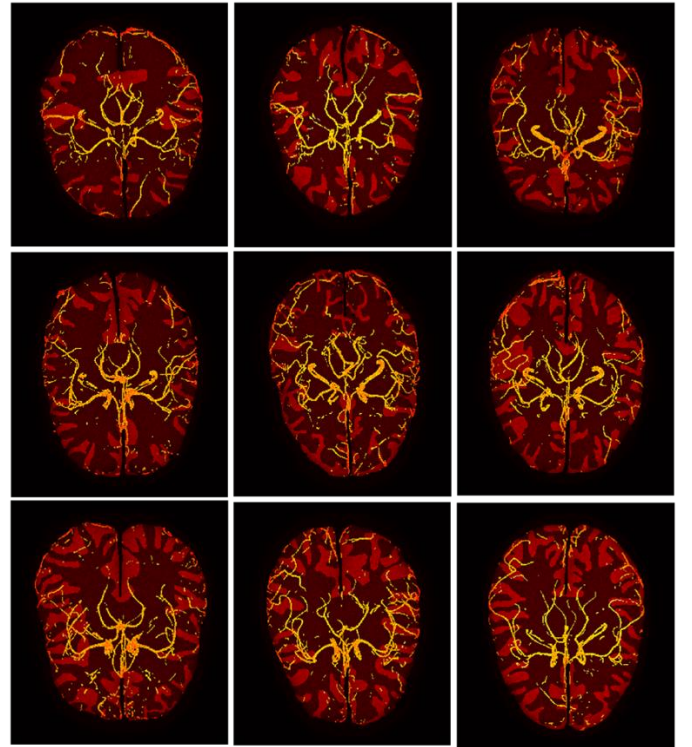

Fig. 5. Nine optical fluence maps of the human brain numerical phantoms

### C. Acoustic Simulation

According to the principle of PA effect, after the pulsed laser illuminates the biological tissue, the absorbed optical energy will be converted into acoustic energy. By multiplying the simulated optical fluence map and the optical absorption coefficients ($\mu_a$), the initial acoustic pressure distribution can be obtained, which is used for the following acoustic simulation.

The acoustic propagation simulation is run using the k-Wave toolbox developed by Bradley Treeby et al [24]. Since we only focus on the optical contrast of the brain tissue, we extract the 12th slice of the optical fluence map as the initial pressure. The k-Wave computational grid is set as follows: 1) the PML size is 40; 2) the number of grid points in the x direction (Nx) and the y direction (Ny) is 320 and 320 respectively; 3) the total grid size in the x direction (x) and y direction (y) is both 150 mm; 4) the grid point spacing in the x direction (dx) and y direction (dy) are both 0.46875 mm. Then the k-Wave computing grid is generated. The acoustic properties of the propagation medium are uniform except for the skull tissue. The sound propagation speed ($c$) in the white matter, gray matter, cerebral cortex, scalp and blood vessel is 1500 m/s, and the sound propagation speed in the skull is 3500 m/s. The second vital parameter is medium density ($d$), which is 1000 g/cm$^3$ in the above five tissues while 1900 g/cm$^3$ in the skull. The ring-shape ultrasound transducer array has 256 elements, whose radius is 74 mm.

The acoustic simulation for each human brain tissue phantom takes about tens of seconds. After obtaining the sensor data, we use the Universal Back-projection algorithm to reconstruct the original image. The result is shown in Fig. 6. From Fig. 6, we could clearly see that the reconstructed image is distorted, which is due to the acoustic aberration of the skull tissue. To obtain ground truth for deep learning-based algorithms, the sound propagation speed and density of the skull tissue are set equal to the other five tissues, and repeat the acoustic simulation process. We obtain the pseudo-ideal reconstructed images and make them as the output data (Ground Truth). The real-scenario aberrated images are set as the input data of the following CNN-based neural network. The input data and output data are used for the following acoustic aberration correction study.

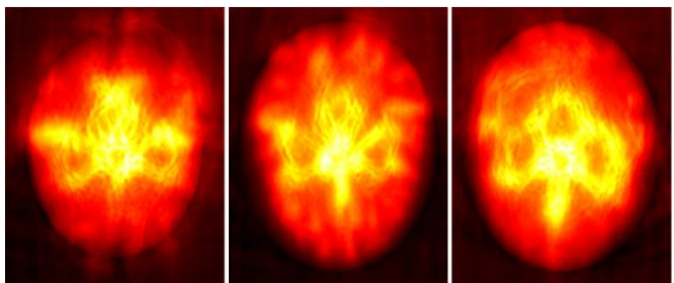

Fig. 6. Three of the reconstructed distorted brain vessel images by Universal back projection method caused by skull aberration.

### D. Acoustic Aberration Correction using U-Net

With the obtained PA sinogram from the acoustic simulation, we want to use deep neural networks to correct the skull acoustic aberration, which is illustrated in Fig. 2 (step 5).

First, we have to pre-process the data and build training and testing datasets. With the acoustic simulation, we can get the paired data (PA sinogram with acoustic aberration and PA sinogram without acoustic aberration). Then we crop and resize the paired sinogram to get input and output data. Their sizes are all 64 x 1200.

Compared with different CNN structure, we think U-net can achieve this task for three reasons. First, the input sinogram is similar in the first half of the data with outputs, and the end of the input sinogram is unwanted reflection aberration signals. So, residual structure is suitable to this work. Interestingly, U-net can be regarded as a big residual structure in some way. Second, the special skip-connection can extract multi-scale features to improve correction outputs. Third, compared with other CNN structures, such as AlexNet, U-net can process images with less training time and computation cost. We illustrated U-net structure in Fig. 7.

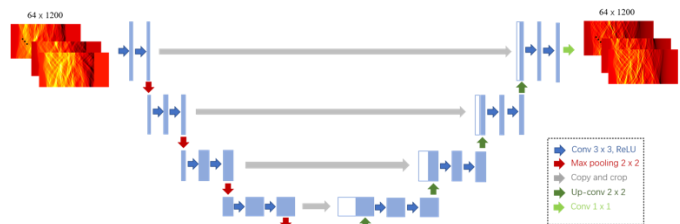

Fig. 7. The structure of U-net used in our paper.

### E. PA Image Reconstruction

We want to use the PA sinogram (with acoustic aberration correction) to reconstruct PA images. Because the U-net has eliminated the skull impact, we can regard the sensor data is generated from homogeneous medium. Then we can apply a lot of classical reconstruction algorithms, such as delay-and-sum (DAS), time reversal, model-based reconstruction and so on. In order to guarantee the reconstruction speed, we implement the DAS algorithm in this work.

## IV. EXPERIMENT AND RESULTS

We have 10 sets of human brain PA sinograms, and we randomly divide them into 7 sets to train, the rest 3 sets to test. The size of each PA sinogram is 256 x 1200. We overlappingly intercept it into 193 PA sub-sinogram (size 64 x 1200). The total training dataset contains 1357 samples. We set the epoch as 50 to avoid the overfitting. We choose ADAM optimizer and the learning rate is 0.001. Mean square error is the loss function. The training process is run with 4 GTX 1080Ti GPU.

The training and testing results are listed in Fig. 8 and Fig. 9. We could see that the corrected sensor data are good enough and similar to the ground truth. And the blood vessel shape distribution could be seen clearly in the reconstructed brain images. The peak signal to noise ratio (PSNR) and structure similarity (SSIM) comparison is shown in Table 2.

Table 2. PSNR and SSIM comparison between Universal back projection and our proposed method

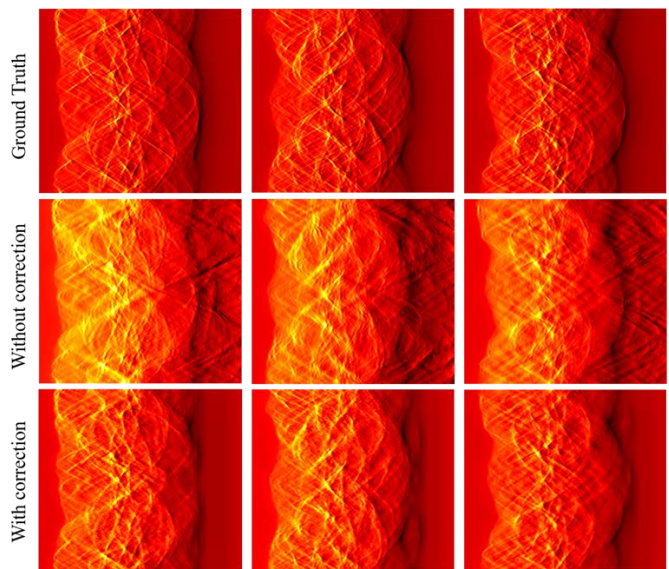

Fig. 8. The partial input (the second row) and output (the first row) sensor data of the deep neural network, and the testing sensor data.

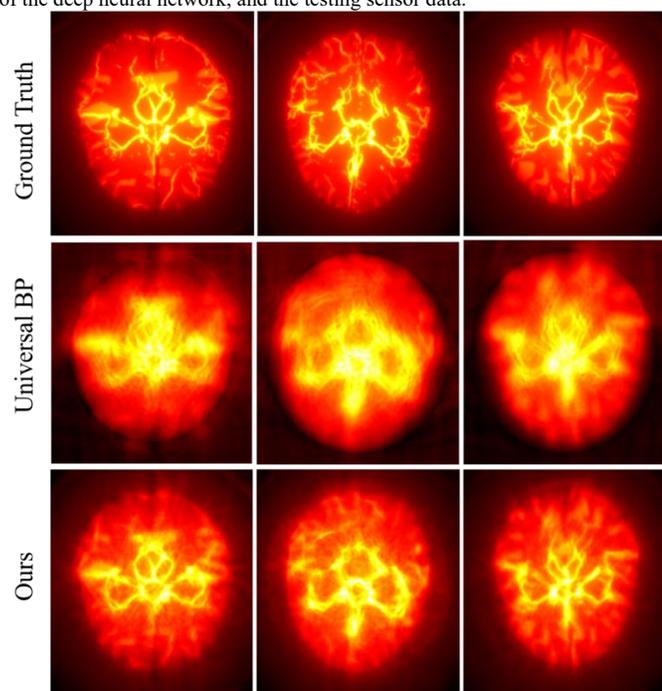

Fig. 9. The respective reconstructed brain images which is obtained by Fig. 8, the first row is the ground truth (output), the second row is the direct Universal BP reconstruction result (input), and the third row is the testing result.

## V. CONCLUSION

In this paper, we proposed to correct the human skull acoustic aberration by using deep learning method for PAT brain imaging. We mainly made 10 human brain numerical phantoms which are translated from MRA modality. Optical and acoustic simulations were performed to simulate the photoacoustic imaging process. In the acoustic simulation part, we made two groups of sensor data. One group takes the inhomogeneous medium and uneven sound speed into consideration and is regarded as the input of the deep neural network (U-net). The other does not consider the heterogeneous property of the medium and is regarded as the output of the deep neural network. The result shows that our method could effectively improve the reconstruction quality and correct the acoustic aberration due to human skull tissue. Deep learning-based method has been proved feasible and effective in correcting acoustic aberration. In the future work, our method will be further validated by ex vivo experiment and in vivo experiment.

## VI. ACKNOWLEDGMENT

This research was funded by Start-up grant of ShanghaiTech University (F-0203-17-004), Natural Science Foundation of Shanghai (18ZR1425000), and National Natural Science Foundation of China (61805139). The authors declare there is no conflict of interest.